\documentclass[slac_one]{revtex4}
\usepackage{graphicx}
\usepackage{fancyhdr}
\newcommand {\Fig}[1]      {Figure~\ref{#1}}
\usepackage{url}
\begin{document}

\title{Jet Analysis in Heavy Ion Collisions in CMS} 

%

\author{M.B.~Tonjes (for the CMS collaboration)}
\affiliation{University of Maryland, College Park, MD, 20742, USA}

\begin{abstract}
    At the Relativistic Heavy Ion Collider, jets have been a useful tool to
    probe the properties of the hot, dense matter created. At the Large Hadron
    Collider, collisions of Pb+Pb at $\sqrt{s_{NN}}$ = 5.5 TeV will provide a
    large cross section of jets at high $E_T$ above the minimum bias heavy ion
    background. Simulations of the Compact Muon Solenoid (CMS) experiment's     
    capability to measure jets in heavy ion collisions are presented. In
    particular, $\gamma$-jet measurements can estimate the amount of energy
    lost by a jet interacting strongly with the medium, since the tagged
    photon passes through unaffected.
\end{abstract}

\maketitle

\thispagestyle{fancy}


\section{INTRODUCTION} 
 In relativistic heavy ion collisions, it is theorized that a hot, dense medium known as the quark-gluon plasma is formed. The energy loss of hard partonic jets propagating through the medium can serve as a useful probe. At the Relativistic Heavy Ion Collider (RHIC), a suppression of hadron yield at high transverse momentum ($p_T$) was observed in 200 GeV Au+Au collisions when compared to p+p collisions (scaled by the number of binary collisions, $N_{coll}$) \cite{WhitePaper}. This jet quenching observation is believed to be from energy loss of fast partons traversing the medium produced. However, high $p_T$ hadrons can also be emitted from the surface of the collision, which makes the quenching measurement harder to quantify \cite{Eskola}. A direct measurement of jet production, as opposed to leading high $p_T$ hadrons, would be illuminating.
 
 At the Large Hadron Collider (LHC), there will be a much higher rate of particle production than at RHIC \cite{Vitev}. In one year of Pb+Pb collisions at  $\sqrt{s_{NN}} = 5.5$ TeV, $0.5$ $nb^{-1}$ integrated luminosity will be obtained, with an estimated 7.8 b inelastic Pb+Pb cross section.  This gives a total of $3.9$ x $10^9$ Pb+Pb collisions. It is expected that fully formed high $E_T$ jets will be made at a rate of more than 10 pairs per second. 

 The Compact Muon Solenoid (CMS) detector is well suited to measure high $p_T$ jets and photons in heavy ion collisions \cite{HIPTDR}. CMS has high precision tracking over $|\eta|<2.5$, calorimetry in $|\eta|<5$, as well as muon identification over $|\eta|<2.5$, with a large bandwidth data acquisition and high level trigger. In events in which a photon and jet are created together, the initial transverse energy of the fragmenting parton can be determined from the photon $E_T$. The modification of the parton fragmentation function by the dense medium in heavy ion collisions can be studied in comparison to p+p collisions.

\section{ANALYSIS DETAILS}

\subsection{Simulations}

For this analysis \cite{GammaJet}, the $\gamma$-jet channel was simulated within a Pb+Pb environment. Simulated events were created by generating p+p events that include high $p_T$ $\gamma$-jet interactions, as well as QCD background. The p+p generators used were PYTHIA \cite{Pythia} and PYQUEN \cite{Pyquen}, where PYQUEN includes a jet quenching scenario. In addition, heavy ion background events were created using HYDJET \cite{Hydjet} at $\sqrt{s_{NN}}$ = 5.5 TeV using either the unquenched or quenched scenario (quenched includes parton energy loss due to the medium). A number of 0-10\% central events equivalent to one year of heavy ion running were created. The $\gamma$-jet signal and Pb+Pb background events were mixed for either the unquenched or quenched scenarios. Data taking conditions were simulated by GEANT-4 with a full CMS detector, followed by full CMS reconstruction.

\subsection{Tracking}

 Charged particle reconstruction was performed using the CMS tracker. The algorithm used is based on seeding from hits in the silicon pixel detector \cite{Tracking}. This algorithm is an extension of that used in p+p collisions with quality cuts optimized for heavy ion collisions. At midrapidity in a heavy ion environment, the algorithmic efficiency is about 70\% near midrapidity for charged particles of $p_T > 1$ GeV/c, with a fake rate of a few percent. For reconstructed tracks of $p_T < 100$ GeV/c, the momentum resolution is $\Delta p_T/p_T$ $<$ 1.5\%.

\subsection{Jet Finding}
Jet finding was performed with the pileup jet finding algorithm using calorimeter energy deposition \cite{PileupJet}. The algorithm is a standard p+p iterative cone jet finding algorithm with a noise/pedestal subtraction. Calorimeter energy from both the electromagnetic (ECAL) and hadronic (HCAL) calorimeters are combined to make towers. Average tower transverse energy and dispersion are calculated for rings in pseudorapidity for each event. Then the tower energy is recalculated by subtracting the mean and dispersion, dropping any towers with negative energy after the subtraction. The first set of jets are found with an iterative cone algorithm (cone radius = 0.5) and the pedestal subtracted tower energy. Then, using the original tower energy, mean and dispersion are found for towers outside of the first set of found jets. This second set of jets have their energy adjusted again with the pedestal subtraction, dropping any negative towers. With these final background subtracted energies, the iterative cone algorithm is used and jets are found. The jet finder has good performance  in heavy ion events, with an efficiency above 80\% for Monte Carlo (MC) jets of $E_T$ above 100 GeV. Reconstructed jets with transverse energy below 30 GeV are cut to reduce the high rate of fakes. Jet energies are not corrected for variations due to pseudorapidity or energy dependent particle response. However, this analysis does not use jet energy except for the $E_T$ cut to reduce the number of fake jets found.
\begin{figure}[hbt]
\begin{center}
\resizebox{0.45\textwidth}{!}{\includegraphics{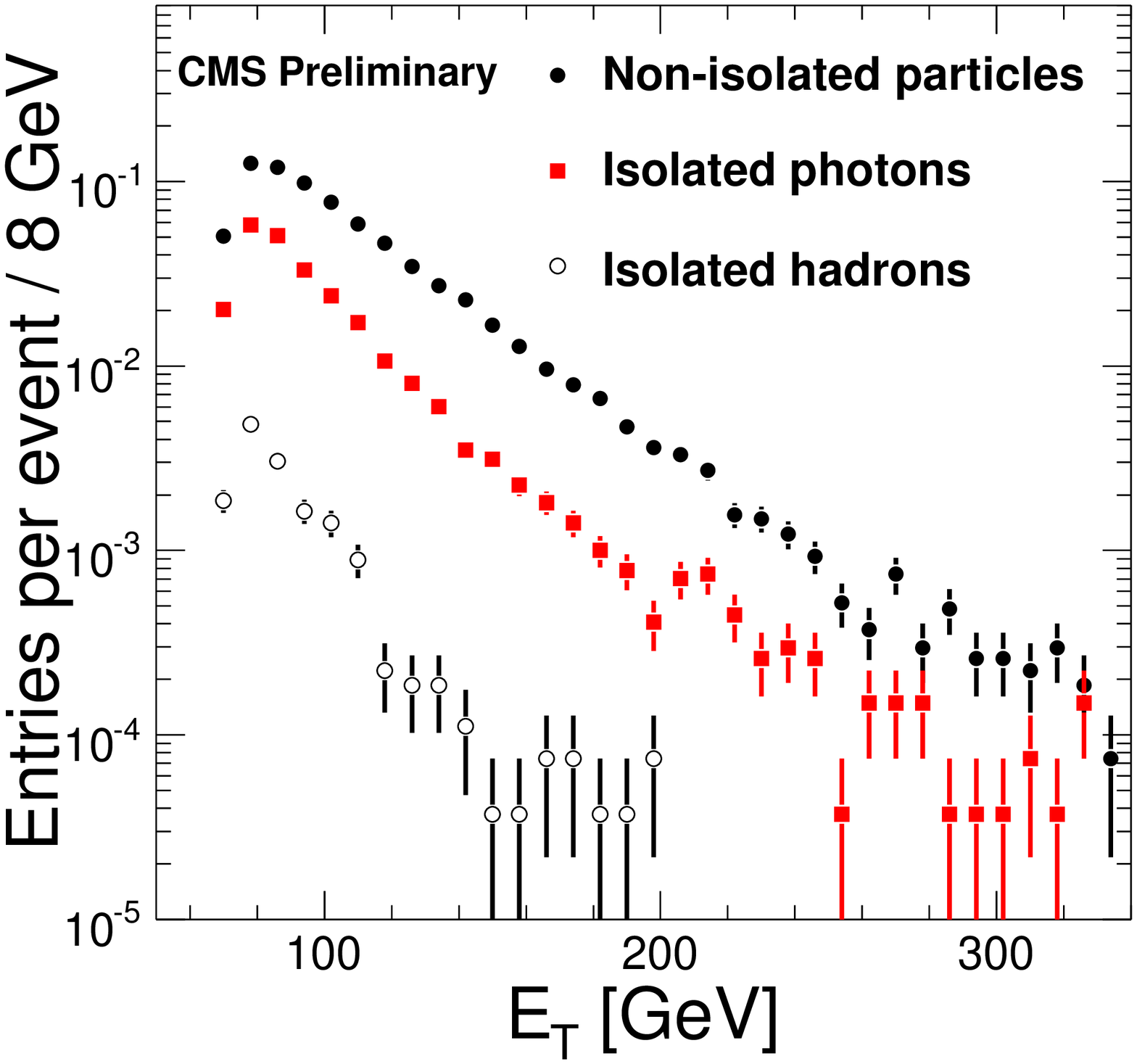}}
\hspace{0.05\textwidth}
\resizebox{0.45\textwidth}{!}{\includegraphics{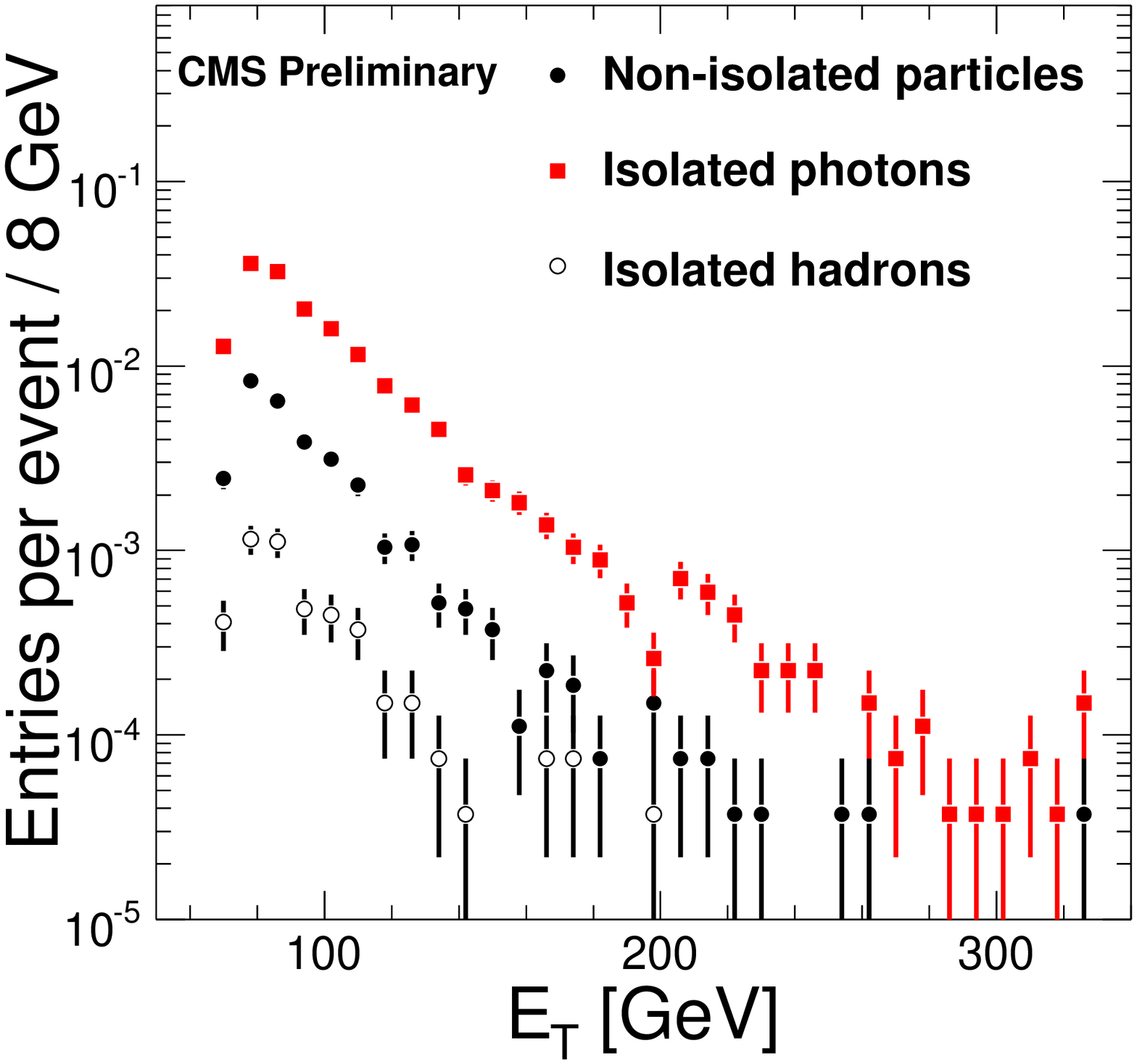}}
\caption{\label{fig:PhotonEff}Transverse energy distribution of photon candidate 
super clusters before application of reconstruction isolation cuts (left panel), and after application of isolation cuts (right panel). The different types of generated
particles: non-isolated particles, isolated photons and isolated hadrons, are classified by applying the 
isolation algorithm to the MC truth. Shown are $0-10$\% central 
quenched Pb+Pb (HYDJET) events.}
\end{center}
\end{figure}
\subsection{Photon Isolation and Multivariate Analysis}
 Photon reconstruction is performed using several parts of the CMS detector. Superclusters of energy deposits in the ECAL that have $E_T > 70$ GeV are found using a standard p+p clustering algorithm. Ten cluster shape variables from the ECAL are combined with isolation variables based on both the ECAL and HCAL, and tracking information. These variables are processed in a multivariate analysis, using the TMVA package of ROOT \cite{TMVA}. The TMVA package is used to determine an optimal cut to divide the candidates into isolated photons (signal) and background. \Fig{fig:PhotonEff} (left panel) shows the $E_T$ distribution comparing isolated photons (red squares) to non-isolated particles (black circles), which has a signal to background ratio of 0.3 before the TMVA analysis. After the TMVA analysis, the signal to background ratio has improved to 4.5, as can be seen in \Fig{fig:PhotonEff} (right panel). For photons above a transverse energy of 70 GeV, the average $E_T$ resolution is about 4.5\%. This analysis includes a fully simulated QCD background with a number of possible false photons: from fragmentation, pions, or even mis-identified photons. However, the signal to background improvement in the heavy ion environment is significant.
  
\section{RESULTS}

\begin{figure}[t]
\begin{center}
\resizebox{0.45\textwidth}{!}{\includegraphics{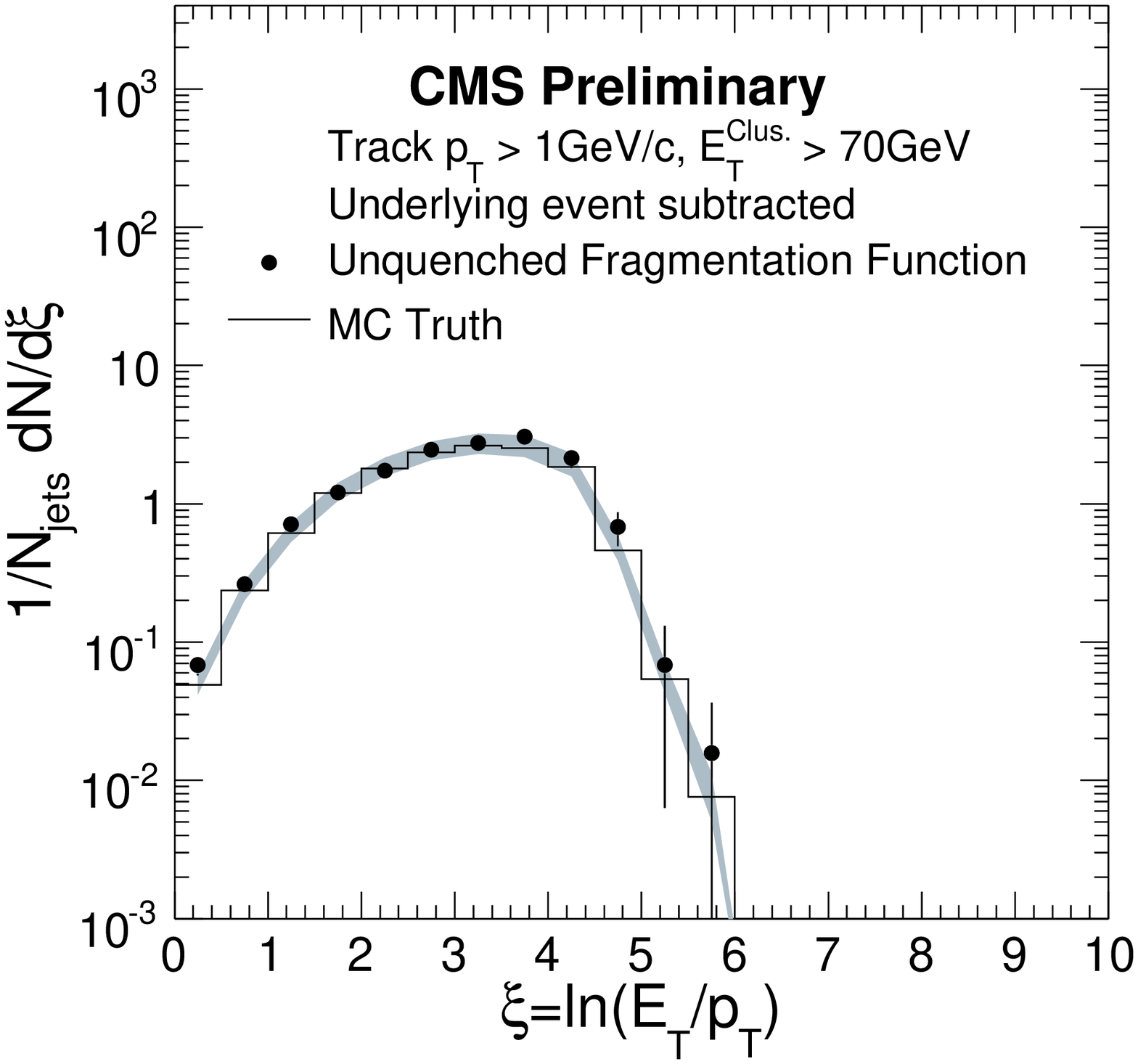}}
\hspace{0.05\textwidth}
\resizebox{0.45\textwidth}{!}{\includegraphics{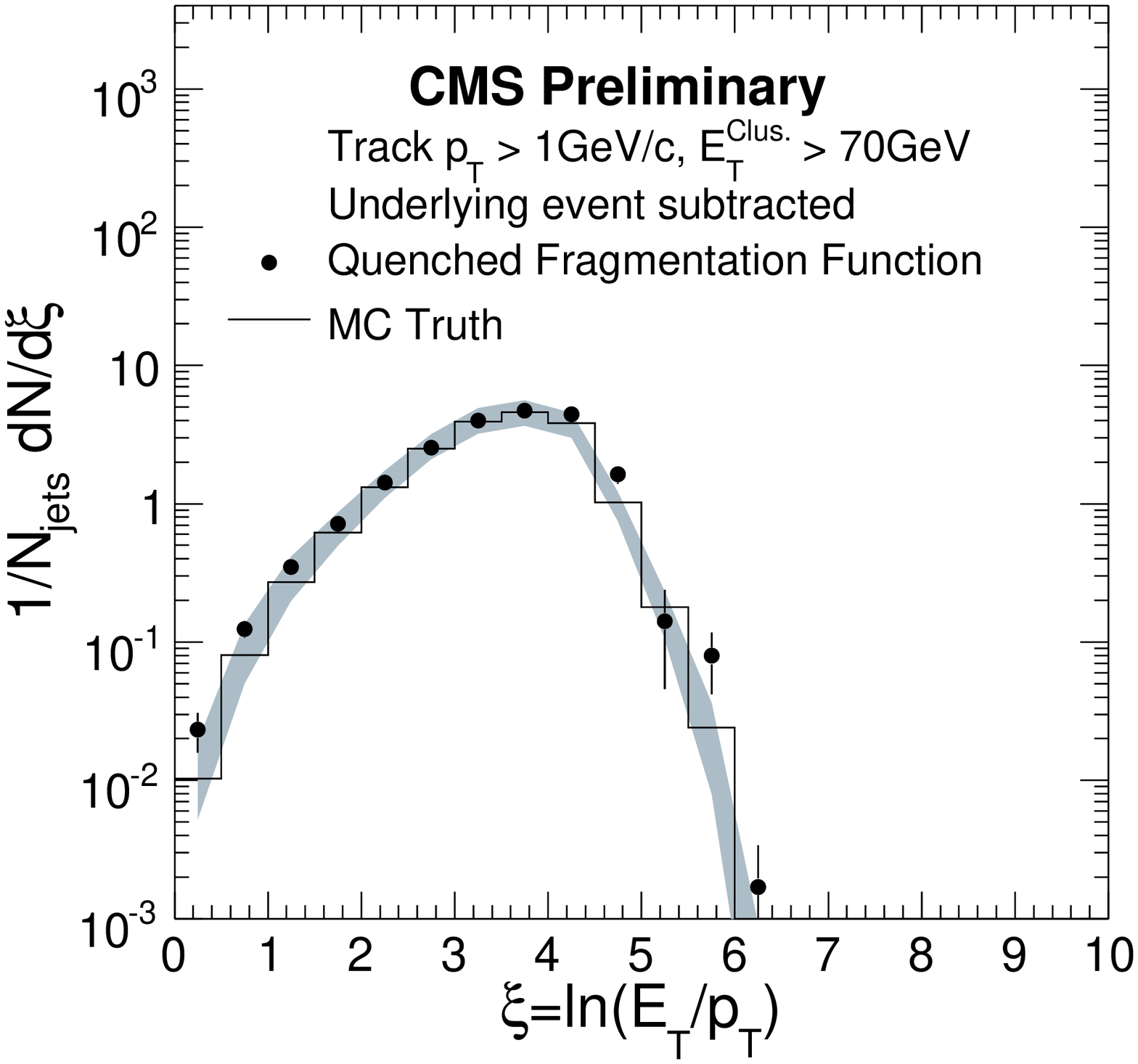}}
\caption{\label{fig:FragFunc}Underlying event subtracted fragmentation functions extracted from central 
Pb+Pb collisions~(symbols) and MC truth (line). Left panel: unquenched $gamma$-jet events (PYTHIA) with unquenched heavy ion background (HYDJET). Right panel: quenched $gamma$-jet events (PYQUEN) with quenched heavy ion background (HYDJET).}
\end{center}
\end{figure}

A fragmentation function is created by correlating isolated photons ($E_T > 70$ GeV) with jets that are back-to-back, that is with an angle of separation between the photon and away side jet greater than $172^{\circ}$. The parton $E_T$ is estimated from the photon $E_T$. The jet $p_T$ is found from reconstructed tracks which are within the jet cone (R=0.5). The variable $\xi$ is then defined as the natural logarithm of the ratio of the photon $E_T$ over the jet $p_T$. The fragmentation function characterizes the process by which high $p_T$ partons become final state hadrons.

To understand the contribution of the underlying heavy ion event, tracks which are outside of the jet cone are studied. The momentum distribution of tracks within a radius of 0.5 cone that is $90^{\circ}$ away from the jet is used to estimate the underlying event contribution. This underlying event fragmentation function is then subtracted from the measured fragmentation function. The results are shown in \Fig{fig:FragFunc} (left panel) for unquenched events, and in \Fig{fig:FragFunc} (right panel) for quenched events. The fragmentation function created from the simulated (MC) truth data before reconstruction is represented by the solid line. There are four main contributions to the systematic errors (grey band) which are added in quadrature. One is QCD jet fragmentation products which pass ECAL cluster cuts and are misidentified as photons. Another is the association of a wrongly paired or fake jet on the away side of the isolated photon. Uncertainty in charged particle reconstruction efficiency also contributes to the final measurement. The largest systematic uncertainty is from the low jet reconstruction efficiency for low $E_T$ jets.

\begin{figure}[htb]
\begin{center}
\resizebox{0.45\textwidth}{!}{\includegraphics{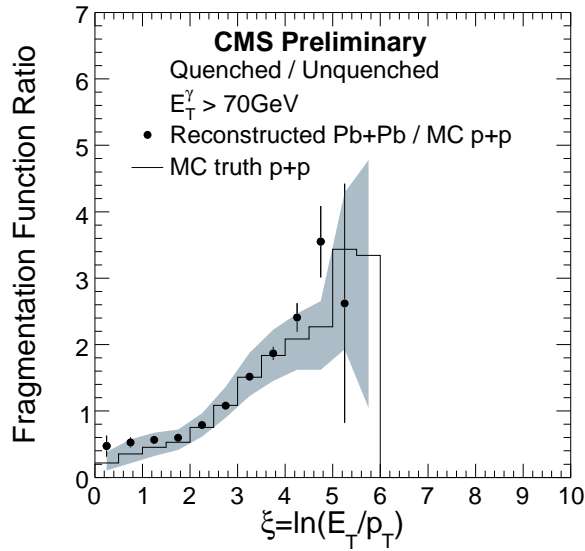}}
\caption{\label{fig:QuenUnquenRatio} Ratio of the reconstructed quenched to unquenched fragmentation functions for events using a minimum $E_T$ for 
the photon candidate ECAL cluster of $70$~GeV~(symbols) and the ratio of the corresponding MC truth fragmentation 
functions~(line).}
\end{center}
\end{figure}

\Fig{fig:QuenUnquenRatio} shows the ratio of the underlying event subtracted fragmentation functions of quenched events to unquenched events, with the systematic errors shown by the grey band. The ratio of the fragmentation functions for simulation truth is shown by the solid line. The agreement between the truth and reconstructed fragmentation functions is quite good. The reconstructed fragmentation function reproduces the MC truth over the full $\xi$ range within uncertainties. The change in the fragmentation function between quenched and unquenched scenarios is larger than the estimated uncertainty.

\section{CONCLUSION}

CMS will be able to quantitatively study high $p_T$ parton fragmentation within the medium by using $\gamma$-jet events. Simulations of one year of Pb+Pb collisions show that this measurement is sensitive to anticipated changes in the fragmentation function within expected statistical and systematic uncertainties.

\end{document}